\documentclass[a4paper, 12pt]{article}
\usepackage{fullpage} 
\usepackage{setspace}
\usepackage{amsmath}
\usepackage{graphicx}
\usepackage{dsfont}
\usepackage{upgreek}
\usepackage{mathtools}
\usepackage{enumitem}

\usepackage[justification=justified,
            format=plain]{caption}
\usepackage[numbers,sort&compress]{natbib}
\allowdisplaybreaks

\usepackage{color}
\usepackage{xcolor}

\usepackage{tabularx}
\usepackage{longtable}

\usepackage{floatrow}
\usepackage{titlesec}
\titleformat*{\section}{\large\bfseries}

\usepackage{lineno}

\renewcommand*{\eqref}[1]{equation~\ref{#1}}

\title{Myelin and saltatory conduction}
\author{
        Maurizio De Pitt\`a \\
            The University of Chicago, USA\\
            EPI BEAGLE, INRIA Rh\^{o}ne-Alpes, France       
        }
\date{\today}

\begin{document}
\maketitle
\vspace{1cm}
\begin{minipage}{\textwidth}
	\centering
	(Submitted as contributed section to the chapter on ``Neurophysiology'' of the book ``\textit{Da\~no cerebral}'' (Brain damage), JC Arango-Asprilla \& L Olabarrieta-Landa eds., Manual Moderno Editions.)  
\end{minipage}

\newpage
\section*{Myelin allows fast and reliable conduction of nerve pulses}
Myelin is a fatty substance that ensheathes the axon of some nerve cells, forming an electrically insulating layer. It is considered a defining characteristic of jawed vertebrates and is essential for the proper functioning of the nervous system of these latter. Myelin is made by different cell types, and varies in chemical composition and configuration, but performs the same insulating function. Myelinated axons look like strings of sausages under a microscope, and because of their white appearance they are integral components of the ``white matter'' of the brain. The myelin sausages ensue from wrapping of axons by myelin in multiple, concentric layers and are separated from each others by small unmyelinated axonal segments known as \textit{nodes of Ranvier}. Typically the length of a node is very small (0.1\%) compared to the length of the myelinated segment. Single myelinated fibers range in diameter from 0.2 to 20~$\upmu$m on average, while unmyelinated fibers range between 0.1 and 1~$\upmu$m \citep{KochB1999}. Peripheral axons are myelinated only if their diameter is larger than about 1~$\upmu$m, and the axonal caliber maintains a rather constant ratio to the myelin sheath thickness \citep{Friede_BR1982}. A normal myelin ensheathing of a mature peripheral nerve is usually 100~times as long as the diameter of the corresponding axon \citep{Friede_JNEN1985}.

Like insulating tape around an electrical wire, myelin helps to insulate the axons from electrically charged atoms and molecules that are found in the fluid surrounding the entire nervous system. Yet, the main purpose of myelin likely is to increase the speed at which neural electrical impulses propagate along the nerve fiber. Along unmyelinated fibers, impulses move continuously as waves, but, in myelinated fibers, they ``hop'' or propagate by a process known as \textit{saltatory conduction}. Myelin in fact decreases capacitance and increases electrical resistance across the cell membrane (the axolemma) thereby helping to prevent the electric current from leaving the axon. This is achieved by a heterogeneous distribution of voltage-gated sodium channels along myelinated fibers, whose density is low ($\sim$25~channels/$\upmu$m$^2$) along myelinated \textit{internodes} (i.e. the myelin ``sausages'') but high at the nodes of Ranvier (between 2000 and 12000~channels/$\upmu$m$^2$) \citep{Saladin_Book2012}. In this fashion, sodium leakage into the extracellular fluid~(ECF) is reduced along the myelinated internodes, thereby maintaining a strong separation of electrical charge between the intracellular fluid and the~ECF. This increases sodium's ability to travel along the axon more freely. However, although sodium diffuses along the axolemma rapidly, this process is decremental by nature so that sodium cannot trigger the opening of the voltage-gated sodium channels as it becomes weaker. The nodes of Ranvier, on the contrary, contain large amounts of voltage-gated sodium channels and are easily excited as they are exposed to the ECF allowing enough sodium into the axon to regenerate the action potential \citep{Brady_Book2011}. An action potential is thus restored to the original depolarization at onset at the axon hillock each time it reaches a node of Ranvier and travels fast along the myelinated axon, hopping from one node of Ranvier to the following one, with a propagation speed that can exceed 10--50~m/s~in humans \citep{Schalow_JANS1995,VanVeen_MM1995}. 

\section*{Specialized glial cells are responsible of myelin production}
Production of myelin -- a process known as \textit{myelination} or \textit{myelogenesis} -- is by specialized glial cells: \textit{oligodendrocytes} myelinate the axons of the central nervous system (CNS), whereas \textit{Schwann cells} supply the myelin for the peripheral nervous system (PNS). Oligodendrocytes and Schwann cells are small cells with relatively few processes, yet with distinct structural differences. While oligodendrocytes can envelop from one to 30~axonal segments (i.e. internodes) depending on axon diameter, one Schwann cell only envelops a single segment of one peripheral axon. Moreover, Schwann cells, but not oligodendrocytes, are surrounded by a basal lamina that forms the demarcation to the mesenchymal environment with the nerve \citep{Martini_2013Ch}. Both oligodendrocytes and Schwann cells not only influence axons by enhancing signal conduction by myelin sheaths but are also responsible of segregating voltage-sensitive ion channels into distinct axonal domains that form the nodes of Ranvier. In addition to formation, maturation and maintenance of nodes of Ranvier are further key functions of these cells \citep{Kandel2012}.

The periodicity of nodes of Ranvier along axons is directly and positively related to axon caliber \citep{Ibrahim_JNS1995}. In the~CNS, internodal lengths are relatively uniform, ranging from 50~to 500~$\upmu$m depending on oligodendrocyte nature \citep{Butt_JNC1998,Murtie_JNR2007}, and nodal periodicity is established developmentally before myelin compaction \citep{Butt_JCN1993,Butt_JNC1997}. As axons grow in length, unmyelinated gaps are myelinated by late developing oligodendrocytes, and this continues well into adulthood, which could explain the extremely short internodes observed in oligodendrocytes of late myelinated areas such as the cortex \citep{Butt_2013Ch}. In the~PNS, on the contrary, the internodal distance can reach 1~to 1.5~mm \citep{Hildebrand_PN1994}. This means that if we consider the human femoral nerve, whose primary axon is approximately 0.5~m, there are approximately 300~to 500~nodes of Ranvier occurring along a primary afferent fiber between the thigh muscle and the cell body in the dorsal root ganglion. Because each internodal segment is formed by a single Schwann cell, then as many as 500~Schwann cells could participate in the myelination of each peripheral sensory axon \citep{Kandel2012}.

Myelinating glial cells and axons are entirely interdependent and should be regarded as functional units rather than separate functional entities per se \citep{Butt_2013Ch}. Neuronal signaling molecules and electrical activity trigger the proliferation of axon-associated glial cells and control the synthesis of myelin \citep{Barres_Nature1993,Demerens_PNAS1996,Wake_Science2011}. Likewise, myelin proteins are essential for axon radial growth and protection from extracellular insults. Additionally, in the presence of injury or pathology, axonal signals are allegedly required for survival of myelinating glial cells, and reciprocally, demyelinated axons do not survive indefinitely when they lose their myelin. Severed myelinated fibers may only regenerate in the PNS but not in the CNS \citep{Kandel2012}.

\section*{Myelination is fundamental for normal brain function}
In humans, myelination begins as early as in the third trimester after birth, although little myelin already exists in the brain at the time of birth -- and quickly occurs during infancy leading to a child's fast development, including crawling and walking in the first year \citep{Yakovlev_1967Ch}. Myelination continues through adolescences up to the fifth decade of life \citep{Bartzokis_JAD2004}. Normal aging, on the other hand, correlates with degenerative changes in myelin, such as shorter internodes or thinners myelin sheaths or splitting, blebbing, ballooning of these latter \citep{Peters_JNC2002,Peters_Glia2008}.

Magnetic resonance imaging has revealed many examples of differences in white matter structure that correlate with specific cognitive abilities such as learning to read \citep{Carreiras_Nature2009}, juggling \citep{Scholz_NN2009} or complex skills like playing the piano \citep{Bengtsson_NN2005}. In this latter case, for example, the level of white matter structure seems to increase proportionately to the number of hours each subject practice piano, indicating that white matter changes in acquiring the skill, rather than performance being pre-determined by a limitation on white matter development. On the other hand, lack of myelination, active \textit{demyelination}, loss of myelinating glial cells or a combination thereof, associate with age-related cognitive decline \citep{Bartzokis_JAD2004,Bowley_JCN2010} as well as neurological disorders like major depressive disorder \citep{Aston_MP2005} and schizophrenia \citep{Takasaki_EJN2010}.
 
When a peripheral fiber is severed, the myelin sheath provides a track along which regrowth can occur. However, the myelin layer does not ensure a perfect regeneration of the nerve fiber. Some regenerated nerve fibers do not find the correct muscle fibers, and some damaged motor neurons of the peripheral nervous system die without regrowth. Damage to the myelin sheath and nerve fiber is often associated with increased functional insufficiency \citep{Kandel2012}.

In general, many diseases of the nervous system involve myelin \citep{Barres_Neuron2008}. Multiple sclerosis~(MS) is one of the most common neurological diseases affecting the~CNS. It involves demyelination, likely due to an autoimmune attack on myelin and oligodendrocytes. Although, in most cases of relapsing and remitting~MS, it appears that there is initial remyelination due to the generation of new oligodendrocytes and new myelin, at some point in the disease this repair process fails leading to unrecoverable motor and cognitive deficits \citep{Lublin_Neurology1996}. In the~PNS, autoimmune reactions to proteins~P$_0$ and~PMP22 that are main components of peripheral myelin produce instead a demyelinating peripheral neuropathy: the Guillain-Barr\'e syndrome \citep{Hughes_TL2005}. Mutations in myelin protein genes also cause a variety of demyelinating diseases in both peripheral and central axons. In the CNS, one example is the Pelizaeus-Merzbacher disease, a leukodystrophy in humans resulting from a recessive mutation of the gene on the long arm of the X-chromosome (Xq21-22) that codes for a myelin protein called proteolipid protein~1 (or~PLP1) \citep{Koeppen_JNEN2002}. In the~PNS, a duplication of the PMP22~gene on chromosome~17 causes instead one form of Charcot-Marie-Tooth disease, which is the most common inherited peripheral neuropathy, and is characterized by progressive muscle weakness, greatly decreased conduction in peripheral nerves, and cycles of demyelination and remyelination \citep{Reilly_JNPNS2011}.

Overall, demyelination slows down, or even stops, conduction of the action potential in an affected axon, because it allows electrical current to leak out of the axonal membrane. Thus, demyelinating diseases have devastating effects on neuronal circuits in the brain, spinal cord, and peripheral nervous system. 

\section*{Iron balance is critical to myelination}
Iron is an essential metal in biological systems where it is necessary for the consumption of oxygen and production of adenosine trisphosphate (ATP) which is a key molecule for intracellular energy transfer. It is also involved in cholesterol and neurotransmitter synthesis and protein degradation. Because of the ability of iron to interact with oxygen, it can also be a potent generator of free radicals. Thus cells have developed an elegant system to keep iron bound to proteins during delivery to cells and storage therein \citep{Connor_AN1992}.

Iron and glia have both distinctive regional and cellular distribution. At the regional level, iron is most prominent in areas associated with motor functions such as basal ganglia, substantia nigra (pars reticulata) and deep cerebellar nuclei and is also present in high concentration in white matter \citep{Connor_Neurosci1990}. At the cellular level, the cells that most prominently need iron for normal function appear to be oligodendrocytes \citep{Connor_Glia1996}. These cells may acquire iron either in earlier stages of development or uptake it as H-ferritin, both from the blood-brain barrier (mainly, but not exclusively, by astrocyte-mediated pathways) and from microglia. Ferritin is then used by oligodendrocytes for maturation and myelin production \citep{Connor_JCN1995}. 

Oligodendrocytes are also the most sensitive to low iron insofar as hypomyelination is a consistent finding of decreased iron availability both in development and adults \citep{Schonberg_JNEN2007}. There is also emerging evidence that iron changes in white matter tracts may mark demyelination, which could be clinically meaningful information to guide timing of treatment intervention \citep{Connor_2013Ch}. A further area of active investigation is on the role of genetic variations that impact brain iron status, and hence myelination, leading to alterations in rates of cognitive decline with age or degree of damage with disease.

Iron balance is critical to glial function given that proinflammatory cytokines, hypoxia and other means of damage to the brain use the ability of iron to generate oxidative stress to promote cell death in all of glial subtypes, but the most sensitive are oligodendrocytes followed by astrocytes and then microglia \citep{Benarroch_Neurol2009}. Remarkably, the ability of iron chelation to protect oligodendrocytes not only from direct exposure to proinflammatory cytokines but also from the toxic effects of activated microglia, suggest that attempts to use this method to treat hypoxic or ischemic injury and demyelinating diseases are well founded \citep{Pedchenko_JNI1998,Sorond_ARS2000}. There is indeed clear evidence that iron chelation will decrease inflammatory reaction, although the challenge remains in the delivery of the iron chelator and the timing of delivery. The chelating compound or the mechanism of delivery must be able to traverse the blood-brain barrier in adequate amounts. Moreover, the chelating compound has to discern between ``good'' and ``bad'' iron not to limit energy production that may be needed for remyelination.

Preliminary experiments in cell culture models suggest that commercially available chelators like deferoxamine~(DFOA) could be viable clinical options to limit damage of glial cells by hypoxia, ischemia and traumatic brain injury, although more research is needed to identify the physiological mechanisms that could regulate the chelating compound both in terms of amount of iron being chelated and the distribution of the chelator with the brain \citep{Robb_BR1998}. Lowering dietary iron may be one way to limit iron available to exacerbate damage associated with neurological trauma, but again, iron must be reintroduced for remyelination and return to normal function \citep{Cortese_ECAP2009,Luders_BP2009}.

Oxidative stress could be a key mechanism in demyelinating disorders. Substantial histological evidence exist indeed for ongoing oxidative stress in demyelination by multiple sclerosis \citep{Levine_BR1997}. Moreover, in support of a role of iron and oxidative stress in promotion of demyelination is the finding that iron deposit in the white matter in multiple sclerosis and that treatment with an iron chelator or an antioxidant can limit the amount of demyelinaton and behavioral abnormalities that are associated with experimental allergic encephalomyelitis \citep{Bowern_JEM1984,Hartung_AN1988,Levine_BR1997}.

A further area that involves oxidative stress damage is radiation used to treat brain tumors. Oligodendrocytes are the most radiosensitive of glial cells \citep{Kim_JNO2008} and human white matter specimens obtained post-radiation confirm a delayed onset of demyelination which correlates with loss of oligodendrocyte function \citep{Panagiotakos_POne2007}. Although there is not direct evidence that iron could contribute to radiosensitivity of oligodendrocytes, it is worth noting that exposing oligodendrocytes in cell cultures to radiation increases oxidative stress sixfold in these cells \citep{Thorburne_JNCHEM1996}, suggesting that such increase in oxidative stress could be prevented by iron chelation.

\newpage
\bibliography{./myelin.bib}

\begin{thebibliography}{52}
\providecommand{\natexlab}[1]{#1}
\providecommand{\url}[1]{\texttt{#1}}
\expandafter\ifx\csname urlstyle\endcsname\relax
  \providecommand{\doi}[1]{doi: #1}\else
  \providecommand{\doi}{doi: \begingroup \urlstyle{rm}\Url}\fi

\bibitem[Koch(1999)]{KochB1999}
C.~Koch.
\newblock \emph{Biophysics of computation: information processing in single
  neurons}.
\newblock Oxford University Press, Inc., New York, 1999.

\bibitem[Friede and Bischhausen(1982)]{Friede_BR1982}
R.~L. Friede and R.~Bischhausen.
\newblock How are sheath dimensions affected by axon caliber and internode
  length?
\newblock \emph{Brain Res.}, 235:\penalty0 335--350, 1982.

\bibitem[Friede and Beuche(1985)]{Friede_JNEN1985}
R.~L. Friede and W.~Beuche.
\newblock A new approach toward analyzing peripheral nerve fiber populations.
  {I. Varance in ssheath thickness corresponds to different geometric
  proportions of the iinternode}.
\newblock \emph{J. Neuropathol. Exp. Neurol.}, 44:\penalty0 60--72, 1985.

\bibitem[Saladin(2012)]{Saladin_Book2012}
K.~S. Saladin.
\newblock \emph{Anatomy \& physiology: the unity of form and function}.
\newblock McGraw-Hill, New York, NY, 6 edition, 2012.

\bibitem[Brady et~al.(2011)Brady, Siegel, {Wayne Albers}, and
  Price]{Brady_Book2011}
S.~T. Brady, G.~J. Siegel, R.~{Wayne Albers}, and D.~L. Price.
\newblock \emph{{Basic neurochemistry: Molecular, cellular and medical
  Aspects}}.
\newblock Lippincott-Raven, Philadelphia, PA, 8 edition, 2011.

\bibitem[Schalow et~al.(1995)Schalow, Zäch, and Warzok]{Schalow_JANS1995}
G.~Schalow, G.~A. Zäch, and R.~Warzok.
\newblock Classification of human peripheral nerve fiber groups by conduction
  velocity and nerve fiber diameter is preserved following spinal cord lesion.
\newblock \emph{J. Aut. Nervous Syst.}, 52:\penalty0 125--150, 1995.

\bibitem[{Van Veen} et~al.(1995){Van Veen}, Schellens, Stegeman, Schoonhoven,
  and {Gabre\"els-Festen}]{VanVeen_MM1995}
B.~K. {Van Veen}, R.~L. L.~A. Schellens, D.~F. Stegeman, R.~Schoonhoven, and
  A.~A. W.~M. {Gabre\"els-Festen}.
\newblock Conduction velocity distributions in normal human sural nerve.
\newblock \emph{Muscle \& Nerve}, 18:\penalty0 1121--1127, 1995.

\bibitem[Martini and Patzk\'o(2013)]{Martini_2013Ch}
R.~Martini and A.~Patzk\'o.
\newblock \emph{Neuroglia}, chapter Schwann cells and myelin, pages 74--85.
\newblock Oxford University Press, 3rd edition, 2013.

\bibitem[Kandel et~al.(2012)Kandel, Schwartz, Jessell, Siegelbaum, and
  Hudspeth]{Kandel2012}
E.~R. Kandel, J.~H. Schwartz, T.~M. Jessell, S.~A. Siegelbaum, and A.~J
  Hudspeth.
\newblock \emph{Principles of Neural Science}.
\newblock McGraw-Hill, New York, NY, 5th edition, 2012.

\bibitem[Ibrahim et~al.(1995)Ibrahim, Butt, and Berry]{Ibrahim_JNS1995}
M.~Ibrahim, A.~M. Butt, and M.~Berry.
\newblock Relationships between myelin sheath diameter and internodal length in
  axons of the anterior medullary velum of the adult rat.
\newblock \emph{J. Neurol. Sci.}, 133\penalty0 (1--2):\penalty0 119--127, 1995.

\bibitem[Butt et~al.(1998)Butt, Ibrahim, and Berry]{Butt_JNC1998}
A.~M. Butt, M.~Ibrahim, and M.~Berry.
\newblock Axon-myelin sheath relations of oligodendrocyte unit phenotypes in
  the adult rat anterior medullary velum.
\newblock \emph{J. Neurocytol.}, 27\penalty0 (4):\penalty0 259--269, 1998.

\bibitem[Murtie et~al.(2007)Murtie, Macklin, and Corfas]{Murtie_JNR2007}
J.~C. Murtie, W.~B. Macklin, and G.~Corfas.
\newblock Morphometric analysis of oligodendrocytes in the adult mouse frontal
  cortex.
\newblock \emph{J. Neurosci. Res.}, 85\penalty0 (10):\penalty0 2080--2086,
  2007.

\bibitem[Butt and Ransom(1993)]{Butt_JCN1993}
A.~M. Butt and B.~R. Ransom.
\newblock Morphology of astrocytes and oligodendrocytes during development in
  the intact rat optic nerve.
\newblock \emph{J. Comp. Neurol.}, 338\penalty0 (1):\penalty0 141--158, 1993.

\bibitem[Butt et~al.(1997)Butt, Ibrahim, and Berry]{Butt_JNC1997}
A.~M. Butt, M.~Ibrahim, and M.~Berry.
\newblock The relationship between developing oligodendrocyte units and
  maturing axons during myelogenesis in the anterior medullary velum of
  neonatal rats.
\newblock \emph{J. Neurocytol.}, 26\penalty0 (5):\penalty0 327--338, 1997.

\bibitem[Butt(2013)]{Butt_2013Ch}
A.~M. Butt.
\newblock \emph{Neuroglia}, chapter Structure and function of oligodendrocytes,
  pages 62--73.
\newblock Oxford University Press, 2013.

\bibitem[Hildebrand et~al.(1994)Hildebrand, Bowe, and
  Remahl]{Hildebrand_PN1994}
C.~Hildebrand, C.~M. Bowe, and I.~N. Remahl.
\newblock Myelination and myelin sheath remodelling in normal and pathological
  {PNS} nerve fibres.
\newblock \emph{Prog. Neurobiol.}, 43:\penalty0 85--141, 1994.

\bibitem[Barres and Raff(1993)]{Barres_Nature1993}
B.~A. Barres and M.~C. Raff.
\newblock Proliferation of oligodendrocyte precurs cells depends on electrical
  activity in axons.
\newblock \emph{Nature}, 361:\penalty0 258--260, 1993.

\bibitem[Demerens et~al.(1996)Demerens, Stankoff, Logak, Anglade, Allinquant,
  Couraud, Zalc, and Lubetzki]{Demerens_PNAS1996}
C.~Demerens, B.~Stankoff, M.~Logak, P.~Anglade, B.~Allinquant, F.~Couraud,
  B.~Zalc, and C.~Lubetzki.
\newblock Induction of myelination in the central nervous system by electrical
  activity.
\newblock \emph{Proc. Natl. Acad, Sci. USA}, 93\penalty0 (18):\penalty0
  9887--9892, 1996.

\bibitem[Wake et~al.(2011)Wake, Lee, and Fields]{Wake_Science2011}
H~Wake, P.~R. Lee, and R.~D. Fields.
\newblock Control of local protein synthesis and initial events in myelination
  by action potentials.
\newblock \emph{Science}, 333\penalty0 (6049):\penalty0 1647--1651, 2011.

\bibitem[Yakovlev and Lecours(1967)]{Yakovlev_1967Ch}
P.~I. Yakovlev and A.~R. Lecours.
\newblock \emph{Regional development of the brain}, chapter The myelogenetic
  cycles of regional maturation of the brain, pages 3--70.
\newblock Blackwell, Oxford, U. K., 1967.

\bibitem[Bartzokis et~al.(2004)Bartzokis, Lu, and Mintz]{Bartzokis_JAD2004}
G.~Bartzokis, P.~H. Lu, and J.~Mintz.
\newblock Quantifying age-related myelin breakdown with {MRI}: novel
  therapeutic targets for preventing cognitive decline and {Alzheimer's}
  disease.
\newblock \emph{J. Alzheimer's Disease}, 6\penalty0 (Suppl. 6):\penalty0
  S53--59, 2004.

\bibitem[Peters(2002)]{Peters_JNC2002}
A.~Peters.
\newblock The effects of normal aging on myelin and nerve fibers: a review.
\newblock \emph{J. Neurocyt.}, 31\penalty0 (8):\penalty0 581--593, 2002.

\bibitem[Peters et~al.(2008)Peters, Verderosa, and Sethares]{Peters_Glia2008}
A.~Peters, A.~Verderosa, and C.~Sethares.
\newblock The neuroglial population in the primary visual cortex of the aging
  rhesus monkey.
\newblock \emph{Glia}, 56\penalty0 (11):\penalty0 1151--1161, 2008.

\bibitem[Carreiras et~al.(2009)Carreiras, Seghier, Baquero, Est{\'e}vez,
  Lozano, Devlin, and Price]{Carreiras_Nature2009}
M.~Carreiras, M.~L. Seghier, S.~Baquero, A.~Est{\'e}vez, A.~Lozano, J.~T.
  Devlin, and C.~J. Price.
\newblock An anatomical signature for literacy.
\newblock \emph{Nature}, 461\penalty0 (7266):\penalty0 983--986, 2009.

\bibitem[Scholz et~al.(2009)Scholz, Klein, Behrens, and
  {Johansen-Berg}]{Scholz_NN2009}
J.~Scholz, M.~C. Klein, T.~E.~J. Behrens, and H.~{Johansen-Berg}.
\newblock Training induces changes in white-matter architecture.
\newblock \emph{Nature Neuroscience}, 12\penalty0 (11):\penalty0 1370--1371,
  2009.

\bibitem[Bengtsson et~al.(2005)Bengtsson, Nagy, Skare, Forsman, Forssberg, and
  Ull{\'e}n]{Bengtsson_NN2005}
S.~L. Bengtsson, Z.~Nagy, S.~Skare, L.~Forsman, H.~Forssberg, and F.~Ull{\'e}n.
\newblock Extensive piano practicing has regionally specific effects on white
  matter development.
\newblock \emph{Nature Neuroscience}, 8\penalty0 (9):\penalty0 1148--1150,
  2005.

\bibitem[Bowley et~al.(2010)Bowley, Cabral, Rosene, and Peters]{Bowley_JCN2010}
M.~P. Bowley, H.~Cabral, D.~L. Rosene, and A.~Peters.
\newblock Age changes in myelinated nerve ffiber of the cingulate bundle and
  corpus callosum in the rhesus monkey.
\newblock \emph{J. Comp. Neurol.}, 518\penalty0 (15):\penalty0 3046--3064,
  2010.

\bibitem[Aston et~al.(2005)Aston, Jiang, and Sokolov]{Aston_MP2005}
C.~Aston, L.~Jiang, and B.~P. Sokolov.
\newblock Transcriptional profiling reveals evidence for signaling and
  oligodendroglial abnormalities in the temporal cortex from patients with
  major depressive disorder.
\newblock \emph{Molecular Psychiatry}, 10\penalty0 (3):\penalty0 309--322,
  2005.

\bibitem[Takasaki et~al.(2010)Takasaki, Yamasaki, Uchigashima, Konno, Yanagawa,
  and Watanabe]{Takasaki_EJN2010}
C.~Takasaki, M.~Yamasaki, M.~Uchigashima, K.~Konno, Y.~Yanagawa, and
  M.~Watanabe.
\newblock Cytochemical and cytological properties of perineuronal
  oligodendrocytes in the mouse cortex.
\newblock \emph{Eur. J. Neurosci.}, 32\penalty0 (8):\penalty0 1326--1336, 2010.

\bibitem[Barres(2008)]{Barres_Neuron2008}
B.A. Barres.
\newblock The mystery and magic of glia: a perspective on their roles in health
  and disease.
\newblock \emph{Neuron}, 60\penalty0 (3):\penalty0 430--440, 2008.

\bibitem[Lublin and Reingold(1996)]{Lublin_Neurology1996}
F.~D. Lublin and S.~C. Reingold.
\newblock Defining the clinical course of multiple sclerosis results of an
  international survey.
\newblock \emph{Neurology}, 46\penalty0 (4):\penalty0 907--911, 1996.

\bibitem[Hughes and Cornblath(2005)]{Hughes_TL2005}
R.~A.~C. Hughes and D.~R. Cornblath.
\newblock {Guillain-Barr\'e} syndrome.
\newblock \emph{The Lancet}, 366\penalty0 (9497):\penalty0 1653--1666, 2005.

\bibitem[Koeppen and Robitaille(2002)]{Koeppen_JNEN2002}
A.~H. Koeppen and Y.~Robitaille.
\newblock {Pelizaeus-Merzbacher} disease.
\newblock \emph{J. Neuropathol. Exp. Neurol.}, 61\penalty0 (9):\penalty0
  747--759, 2002.

\bibitem[Reilly et~al.(2011)Reilly, Murphy, and Laur{\'a}]{Reilly_JNPNS2011}
M.~M. Reilly, S.~M. Murphy, and M.~Laur{\'a}.
\newblock {Charcot-Marie-Tooth} disease.
\newblock \emph{J. Periph. Nervous Sys.}, 16\penalty0 (1):\penalty0 1--14,
  2011.

\bibitem[Connor and Benkovic(1992)]{Connor_AN1992}
J.~R. Connor and S.~A. Benkovic.
\newblock Iron regulation in the brain: histochemical, biochemical, and
  molecular considerations.
\newblock \emph{Ann. Neurol.}, 32\penalty0 (S1), 1992.

\bibitem[Connor and Menzies(1990)]{Connor_Neurosci1990}
J.~R. Connor and S.~L. Menzies.
\newblock Altered cellular distribution of iron in the central nervous system
  of myelin deficient rats.
\newblock \emph{Neuroscience}, 34\penalty0 (1):\penalty0 265--271, 1990.

\bibitem[Connor and Menzies(1996)]{Connor_Glia1996}
J.~R. Connor and S.~L. Menzies.
\newblock Relationship of iron to oligondendrocytes and myelination.
\newblock \emph{Glia}, 17\penalty0 (2):\penalty0 83--93, 1996.

\bibitem[Connor et~al.(1995)Connor, Pavlick, Karli, Menzies, and
  Palmer]{Connor_JCN1995}
J.~R Connor, G.~Pavlick, D.~Karli, S.~L. Menzies, and C.~Palmer.
\newblock A histochemical study of iron-positive cells in the developing rat
  brain.
\newblock \emph{J. Comp. Neurol.}, 355\penalty0 (1):\penalty0 111--123, 1995.

\bibitem[Schonberg et~al.(2007)Schonberg, Popovich, and
  {McTigue}]{Schonberg_JNEN2007}
D.~L. Schonberg, P.~G. Popovich, and D.~M. {McTigue}.
\newblock Oligodendrocyte generation is differentially influenced by toll-like
  receptor {(TLR)~2 and TLR4-mediated} intraspinal macrophage activation.
\newblock \emph{J. Neuropathol. Exp. Neurol.}, 66\penalty0 (12):\penalty0
  1124--1135, 2007.

\bibitem[Connor(2013)]{Connor_2013Ch}
J.~R. Connor.
\newblock \emph{Neuroglia}, chapter Iron and glia, pages 586--602.
\newblock Oxford University Press, 2013.

\bibitem[Benarroch(2009)]{Benarroch_Neurol2009}
E.~E. Benarroch.
\newblock Oligodendrocytes: susceptibility to injury and involvement in
  neurologic disease.
\newblock \emph{Neurology}, 72\penalty0 (20):\penalty0 1779--1785, 2009.

\bibitem[Pedchenko and {LeVine}(1998)]{Pedchenko_JNI1998}
T.~V. Pedchenko and S.~M. {LeVine}.
\newblock Desferrioxamine suppresses experimental allergic encephalomyelitis
  induced by {MBP in SJL} mice.
\newblock \emph{J. Neuroimmunol.}, 84\penalty0 (2):\penalty0 188--197, 1998.

\bibitem[Sorond and Ratan(2000)]{Sorond_ARS2000}
F.~A. Sorond and R.~R. Ratan.
\newblock Ironing-out mechanisms of neuronal injury under hypoxic--ischemic
  conditions and potential role of iron chelators as neuroprotective agents.
\newblock \emph{Antioxid. Redox Signal.}, 2\penalty0 (3):\penalty0 421--436,
  2000.

\bibitem[Robb and Connor(1998)]{Robb_BR1998}
S.~J. Robb and J.~R. Connor.
\newblock An in vitro model for analysis of oxidative death in primary mouse
  astrocytes.
\newblock \emph{Brain Res.}, 788\penalty0 (1):\penalty0 125--132, 1998.

\bibitem[Cortese et~al.(2009)Cortese, Konofal, Bernardina, Mouren, and
  Lecendreux]{Cortese_ECAP2009}
S.~Cortese, E.~Konofal, B.~D. Bernardina, M.-C. Mouren, and M.~Lecendreux.
\newblock Sleep disturbances and serum ferritin levels in children with
  attention-deficit/hyperactivity disorder.
\newblock \emph{Eur. Child Adolesc. Psychiatry}, 18\penalty0 (7):\penalty0
  393--399, 2009.

\bibitem[Luders et~al.(2009)Luders, Narr, Hamilton, Phillips, Thompson, Valle,
  {Del'Homme}, Strickland, {McCracken}, Toga, and Levitt]{Luders_BP2009}
E.~Luders, K.~L. Narr, L.~S. Hamilton, O.~R. Phillips, P.~M. Thompson, J.~S.
  Valle, M.~{Del'Homme}, T.~Strickland, J.~T. {McCracken}, A.~W. Toga, and
  J.~G. Levitt.
\newblock Decreased callosal thickness in attention-deficit/hyperactivity
  disorder.
\newblock \emph{Biol. Psychiatry}, 65\penalty0 (1):\penalty0 84--88, 2009.

\bibitem[{LeVine}(1997)]{Levine_BR1997}
S.~M. {LeVine}.
\newblock Iron deposits in multiple sclerosis and {Alzheimer's} disease brains.
\newblock \emph{Brain Res.}, 760\penalty0 (1):\penalty0 298--303, 1997.

\bibitem[Bowern et~al.(1984)Bowern, Ramshaw, Clark, and
  Doherty]{Bowern_JEM1984}
N.~Bowern, I.~A. Ramshaw, I.~A. Clark, and P.~C. Doherty.
\newblock Inhibition of autoimmune neuropathological process by treatment with
  an iron-chelating agent.
\newblock \emph{J. Exp. Med.}, 160\penalty0 (5):\penalty0 1532--1543, 1984.

\bibitem[Hartung et~al.(1988)Hartung, Sch{\"a}fer, Heininger, and
  Toyka]{Hartung_AN1988}
H.-P. Hartung, B.~Sch{\"a}fer, K.~Heininger, and K.~V. Toyka.
\newblock Suppression of experimental autoimmune neuritis by the oxygen radical
  scavengers superoxide dismutase and catalase.
\newblock \emph{Annals Neurol.}, 23\penalty0 (5):\penalty0 453--460, 1988.

\bibitem[Kim et~al.(2008)Kim, Brown, Jenrow, and Ryu]{Kim_JNO2008}
J.~H. Kim, S.~L. Brown, K.~A. Jenrow, and S.~Ryu.
\newblock Mechanisms of radiation-induced brain toxicity and implications for
  future clinical trials.
\newblock \emph{J. Neurooncol.}, 87\penalty0 (3):\penalty0 279--286, 2008.

\bibitem[Panagiotakos et~al.(2007)Panagiotakos, Alshamy, Chan, Abrams,
  Greenberg, Saxena, Bradbury, Edgar, Gutin, and Tabar]{Panagiotakos_POne2007}
G.~Panagiotakos, G.~Alshamy, B.~Chan, R.~Abrams, E.~Greenberg, A.~Saxena,
  M.~Bradbury, M.~Edgar, P.~Gutin, and V.~Tabar.
\newblock Long-term impact of radiation on the stem cell and oligodendrocyte
  precursors in the brain.
\newblock \emph{PloS One}, 2\penalty0 (7):\penalty0 e588, 2007.

\bibitem[Thorburne and Juurlink(1996)]{Thorburne_JNCHEM1996}
S.~K. Thorburne and B.~H.~J. Juurlink.
\newblock Low glutathione and high iron govern the susceptibility of
  oligodendroglial precursors to oxidative stress.
\newblock \emph{J. Neurochem.}, 67\penalty0 (3):\penalty0 1014--1022, 1996.

\end{thebibliography}

\end{document}